\documentclass[12pt]{iopart}
\usepackage{times}
\usepackage{graphicx}

\begin{document}

\title{Excitonic complexes in quantum Hall systems}

\author{A W\'ojs and J J Quinn}
\address{University of Tennessee, Knoxville}

\begin{abstract}
The formation and various possible decay processes of neutral 
and charged excitonic complexes in electronic integral and 
fractional quantum Hall systems are discussed.
The excitonic complexes are bound states of a small number of 
the relevant negatively and positively charged quasiparticles 
(e.g., conduction electrons and valence holes, reversed-spin 
electrons and spin holes, Laughlin quasielectrons and quasiholes, 
composite fermions) that occur in an electron system under 
specific conditions (e.g., electron density, well width, 
electric and magnetic fields, or hydrostatic pressure).
The examples of such bound states are interband neutral and 
charged excitons, fractionally charged ``anyon excitons,'' 
spin waves, skyrmions, or ``skyrmion excitons.''
Their possible decay processes include radiative recombination,
experimentally observed in photoluminescence or far infrared
emission, or spin transitions, important in the context of 
nuclear spin relaxation.
\end{abstract}

\section{Introduction}

The transport, optical, and spin properties of a two-dimensional 
electron gas (2DEG) in a high magnetic field have been intensively 
studied both experimentally \cite{Klitzing80,Tsui82,Heiman88,Kheng93,%
Barret94,Tycko95} and theoretically \cite{Laughlin81,Laughlin83,%
Haldane83,Kallin84,Rezayi87,Vagner88,Jain89,MacDonald90,MacDonald92,%
Sondhi93} over more than a decade.
Some of these studies have demonstrated that a common scenario of 
the formation of what can generally be called an excitonic complex 
occurs in various seemingly different physical situations.
The excitonic complexes, consisting of a small number of appropriate 
elementary charged excitations (positively and negatively charged 
quasiparticles of various type depending on a particular form of 
the electron--electron correlations in the underlying 2DEG), can 
often be considered as nearly free particles with well defined
single-particle properties.
These properties, such as electric charge, characteristic size, 
longitudinal or angular momentum, spin, binding energy, or oscillator 
strength for a particular type of quasiparticle--antiquasiparticle 
recombination process, determine the response of the 2DEG to the 
experimental perturbation.
In particular, being weakly coupled to one another or to the 
electrons, excitonic complexes recombine obeying simple selection 
rules that result from their geometric (2D translational) or 
dynamical (particle--hole) symmetries.
These simple symmetries often persist under experimental conditions
despite complicated electron--electron correlations or such
typical symmetry-breaking mechanisms as disorder or collisions, 
and greatly simplify the measured response of the entire system.
Sometimes, such simplification is even undesirable as it can make
the experiment sensitive only to the simple properties of the 
excitonic complexes, and quite insensitive to the specific properties 
of the underlying 2DEG.

For example, it has long been predicted that the photoluminescence 
(PL) spectrum in an infinitely high magnetic field contains no 
information about the electron--electron correlations (e.g., the 
presence or charge of Laughlin quasiparticles in the fractional 
quantum Hall regime) regardless of possible disorder \cite{MacDonald92}.
Instead, the spectrum is reduced to a single discrete transition 
corresponding to the recombination of a neutral exciton in the zero 
momentum ground state, and either decreasing the magnetic field 
in order to allow interactions to admix higher Landau levels (LL's) 
or applying an electric field to spatially separate electrons and 
holes is needed for PL to become a useful tool for studying 
electron--electron interactions.

Another example is related to a prediction \cite{x-td,Palacios96} 
that the most strongly bound complex involving conduction electrons 
($e$) and a valence hole ($v$) in very high magnetic fields is a 
triplet state of the charged exciton ($X^-=2e+v$).
This state is nonradiative because of both geometrical and dynamical 
symmetry, and has not been experimentally confirmed in earlier 
experiments in high magnetic fields \cite{Hayne99}, but only quite 
recently \cite{Yusa01,Schuller02}, when special measures were taken 
to detect its weak PL signal.
While breaking of the dynamical, particle--hole symmetry in a finite
magnetic field is by no means surprising, the fact that collisions
of an $X^-$ with the surrounding electrons do not relax the geometrical
selection rule associated with the angular momentum conservation 
is a nice demonstration of Laughlin correlations of the $X^-$ with 
other negative charges \cite{x-fqhe,x-cf}.
As a result of these correlations, at small values of the filling
factor $\nu$, the $X^-$'s remain spatially isolated and avoid high 
energy collisions with one other or with electrons to become true 
quasiparticles of a 2DEG containing additional valence holes \cite{x-tb}.

In the following sections of this article we will review a few
examples of excitonic complexes that form in electronic quantum 
Hall systems:
interband excitonic complexes in Sec.~\ref{sec_X}, anyon excitons 
in Sec.~\ref{sec_AX}, skyrmions in Sec.~\ref{sec_Sky}, and skyrmion 
excitons in Sec.~\ref{sec_SkyX}.
We will discuss the similarities and differences between all 
these complexes, and show the role they play in experimental 
studies of the 2DEG, particularly in PL.

\section{Model}

The numerical results presented here are obtained by exact 
numerical diagonalization of the interaction Hamiltonian of 
a finite number $N$ of electrons (and, sometimes, one or more
valence holes) confined on a spherical surface of radius $R$.
In this model, the radial magnetic field $B$ is due to a monopole 
placed in the center of the sphere \cite{Haldane83}.
The monopole strength $2Q$ is defined in the units of elementary 
flux $\phi_0=hc/e$, so that $4\pi R^2B=2Q\phi_0$ and the magnetic 
length is $\lambda=R/\sqrt{Q}$.
The single-particle states are the eigenstates of angular momentum 
$l$ and its projection $m$ and are called monopole harmonics.
The energies $\varepsilon$ fall into $(2l+1)$-fold degenerate angular 
momentum shells separated by the cyclotron energy $\hbar\omega_c$.
The $n$-th ($n\ge0$) shell (LL) has $l=Q+n$ and thus $2Q$ is a measure 
of the system size through the LL degeneracy.
Due to the spin degeneracy, each $l$-shell is further split by the 
Zeeman gap, $E_{\rm Z}$.

Using a composite index $i=[nm\sigma]$ ($\sigma$ is the spin 
projection), the Hamiltonian of interacting particles can be 
written as $H=\sum c_{i\alpha}^\dagger c_{i\alpha}\varepsilon_{i\alpha}
+\sum c_{i\alpha}^\dagger c_{j\beta}^\dagger c_{k\beta} c_{l\alpha} 
V_{ijkl\alpha\beta}$, where $c_{i\alpha}^\dagger$ and $c_{i\alpha}$ 
create and annihilate particle $\alpha$ (conduction electron $e$ or 
valence hole $v$, reversed-spin electron $e_{\rm R}$ or spin hole 
$h$, etc.) in state $i$ with energy $\varepsilon_{i\alpha}$, and 
$V_{ijkl\alpha\beta}$ are the interaction (Coulomb) matrix elements.
Hamiltonian $H$ is diagonalized in the basis of Slater determinants.
The result of the diagonalization procedure is the set of many-body 
eigenenergies and eigenvectors.
The energies $E$ will be shown as a function of the conserved orbital 
($L$ and $L_z$) and spin ($S$ and $S_z$) quantum numbers.
To interpret the results obtained in the spherical geometry for the 
infinite planar system, $L$ and $L_z$ must be appropriately translated
into the corresponding planar quantities \cite{x-tb,sky}.
For example, for the (charge or spin) wave eigenstates that carry no
net charge, angular momentum $L$ must be replaced by wave vector $k=L/R$, 
while for the eigenstates corresponding to charged excitations $L$ and 
$L_z$ are connected with planar angular momentum projection ${\cal M}$ 
and its center-of-mass component ${\cal M}_{\rm CM}$. 
The eigenvectors $\left|\psi\right>$ are needed to calculate spectral 
functions to describe PL or other decay processes, 
$\tau_{if}^{-1}=|\left<f\right|{\cal P}\left|i\right>|^2$, where 
$\psi=i$ or $f$ are the initial and final states, respectively, and 
${\cal P}$ is the appropriate transition operator.

\section{Neutral and charged interband excitons}\label{sec_X}

\begin{figure}
\begin{center}
\includegraphics{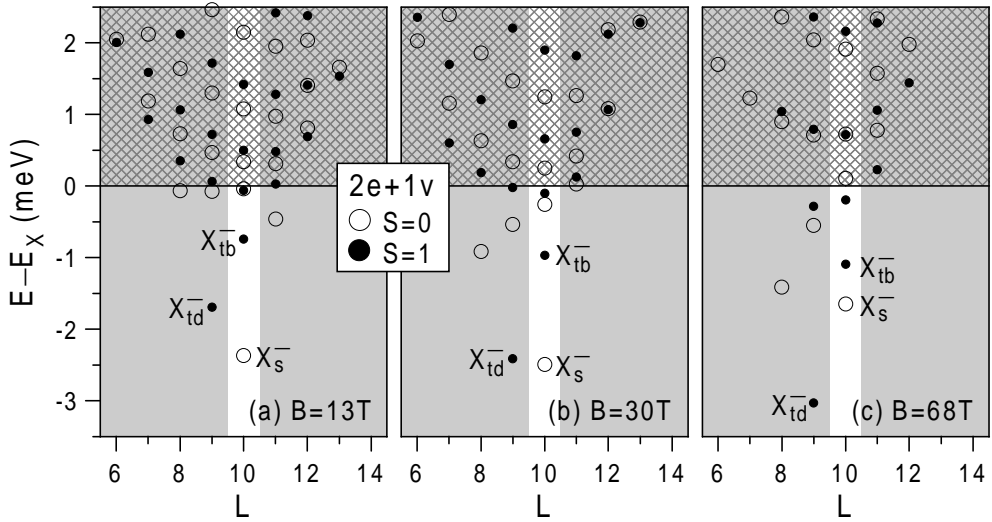}
\end{center}
\caption{
   The energy spectra (energy $E$ vs.\ angular momentum $L$) of 
   the $2e+v$ system in a symmetric GaAs quantum well of width 
   $w=11.5$~nm at the magnetic field $B=13$~T (a), 30~T (b), and 
   68~T (c), calculated on Haldane sphere with LL degeneracy 
   $2Q+1=21$.}
\label{fig01}
\end{figure}

An $X^-=2e+v$ consists of only three particles.
The energy spectra of this simple system are shown in Fig.~\ref{fig01} 
for a GaAs symmetric quantum well of width $w=11.5$~nm and for 
$B=13$, 30, and 68~T.
The effects of LL mixing, finite well width, anisotropy of the hole 
mass and its dependence on $B$, and the realistic Zeeman gap 
$E_{\rm Z}$ have all been included \cite{x-tb}. 
The energy $E$ is measured from the exciton energy $E_{\rm X}$, 
so that for the bound $X^-$ states it gives the binding energy 
$\Delta=E_{\rm X}-E$, and both singlet and triplet electron spin 
configurations are shown.

\begin{figure}
\begin{center}
\includegraphics{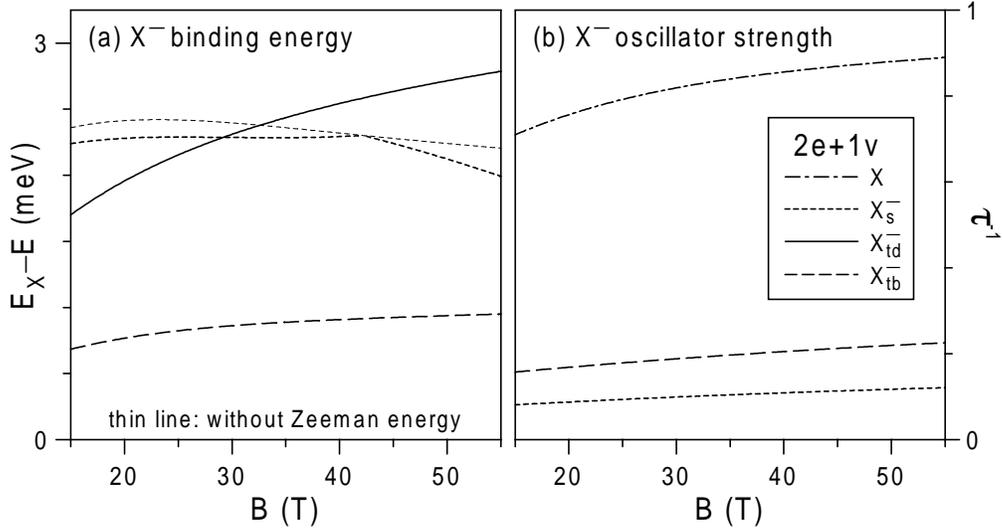}
\end{center}
\caption{
   The binding energies $\Delta$ (a) and oscillator strengths 
   $\tau^{-1}$ (b) of different $X^-$ states in a symmetric GaAs 
   quantum well of width $w=11.5$~nm, as a function of magnetic 
   field $B$.}
\label{fig02}
\end{figure}

Because the emission of a photon does not change angular momentum
of the (envelope) electron wave function, and because the electron
left in the lowest LL after the radiative $X^-$ recombination has
$l=Q$, only those $X^-$ states at $L=Q$ are optically active. 
Of all bound $X^-$ states in Fig.~\ref{fig01}, three are of 
particular importance.
The $X^-_{\rm s}$ (singlet) and $X^-_{\rm tb}$ (triplet-bright) 
are the only strongly bound radiative states, while $X^-_{\rm td}$ 
(triplet-dark) has by far the lowest energy of all non-radiative 
states.
The relative energy of different $X^-$ states depends on 
experimentally variable parameters (e.g., $B$, $w$, or $E_{\rm Z}$),
and indeed, the transition between the $X^-_{\rm s}$ and $X^-_{\rm td}$ 
states can be seen in Fig.~\ref{fig01}(b).
The binding energies $\Delta$ and oscillator strengths $\tau^{-1}$ 
of the three $X^-$ states, extrapolated to the $R/\lambda=\sqrt{Q}
\rightarrow\infty$ limit, have been plotted in Fig.~\ref{fig02} as 
a function of $B$.
The $X^-_{\rm s}\leftrightarrow X^-_{\rm td}$ transition is found 
at $B\approx30$~T, and the $X^-_{\rm tb}$ state is about two times 
``brighter'' than $X^-_{\rm s}$ (although both are considerably 
``darker'' than the $X$).

\begin{figure}
\begin{center}
\includegraphics{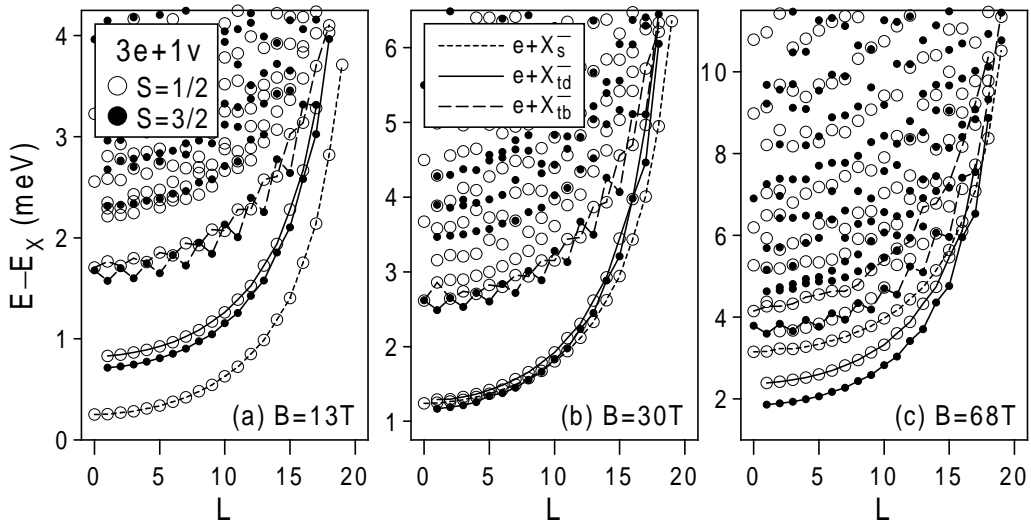}
\end{center}
\caption{
   The same as in Fig.~\ref{fig01} but for the $3e+v$ system.}
\label{fig03}
\end{figure}

Even in dilute systems, one might expect that collisions with 
surrounding electrons can affect the $X^-$ recombination and 
in particular allow for weak emission from $X^-_{\rm td}$.
The surprising experimental fact that the effect of such collisions 
is minimal \cite{Hayne99,Yusa01,Schuller02} results from Laughlin 
correlations between $X^-$ and electrons in the fractional quantum 
Hall regime \cite{x-fqhe,x-cf}.
In Fig.~\ref{fig03} we plot the energy spectra of $3e+v$ systems,
in which the lowest bands of states describe repulsion of different 
$e$--$X^-$ pairs.
The dependence of pair interaction energy $V$ on pair angular 
momentum $L$ is the interaction pseudopotential, which completely 
determines correlations in a degenerate LL.
It is known that if $V(L)$ is ``superharmonic'' ($V$ decreases 
more quickly than linearly as a function of separation $\left<
r^2\right>$ when $L$ is decreased), then Laughlin correlations 
occur \cite{parent}.
It turns out that $e$--$X^-$ pseudopotential is superharmonic
(similar to the $e$--$e$ pseudopotential in the lowest LL). 
The resulting Laughlin correlations between an $X^-$ and the 
electrons mean that one or more $e$--$X^-$ pair states of highest 
repulsion are maximally avoided, or in other words, that the high 
energy $e$--$X^-$ collisions do not occur.

\begin{figure}
\begin{center}
\includegraphics{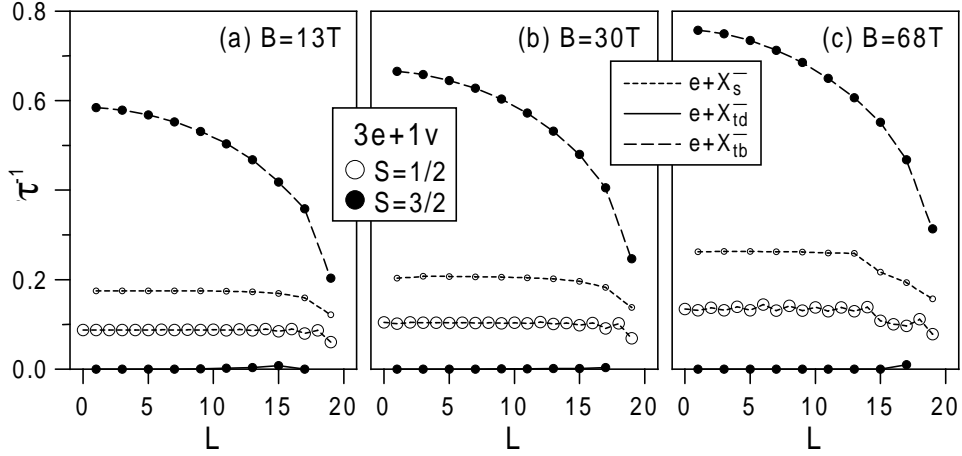}
\end{center}
\caption{
   The oscillator strengths $\tau^{-1}$ of different $X^-$ states 
   interacting with an electron in a symmetric GaAs quantum well 
   of width $w=11.5$~nm at the magnetic field $B=13$~T (a), 30~T (b), 
   and 68~T (c), calculated on Haldane sphere with LL degeneracy 
   $2Q+1=21$, and plotted as a function of the $e$--$X^-$ pair 
   angular momentum $L$.}
\label{fig04}
\end{figure}

\begin{figure}
\begin{center}
\includegraphics{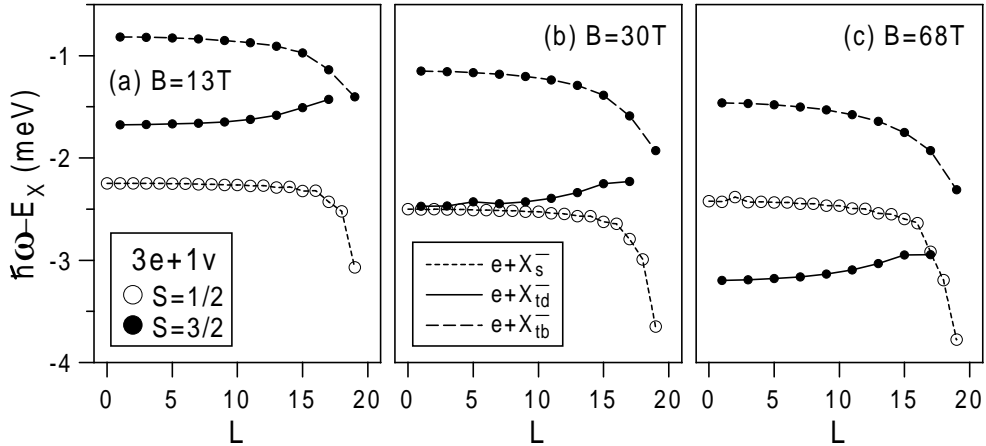}
\end{center}
\caption{
   The same as in Fig.~\ref{fig04} but with recombination 
   energy $\hbar\omega$ shown on vertical axes.}
\label{fig05}
\end{figure}

In Figs.~\ref{fig04} and \ref{fig05} we plot the oscillator strengths 
$\tau^{-1}$ and emission energies $\hbar\omega$ for the $3e+v$ 
eigenstates corresponding to an $X^-$ interacting with an electron.
In both figures, the horizontal axes give pair angular momentum $L$ 
which in a Laughlin correlated system is simply related to the LL 
filling factor $\nu$ (only the $L\le l_{X^-}+l_e-\mu$ pair states 
occur at $\nu\le\mu^{-1}$).
As expected, for small $L$ (i.e., very dilute 2DEG) both $\hbar\omega$ 
and $\tau^{-1}$ converge to the values appropriate for single $X^-$'s 
plotted in Fig.~\ref{fig02}, meaning that there is no significant 
effect of the $e$--$X^-$ interactions on the $X^-$ recombination at 
small $\nu$.
Somewhat surprisingly, the Laughlin correlations prevent considerable 
increase of the $\tau^{-1}_{\rm td}$ through interaction with electrons 
even at $\nu\approx{1\over3}$.
This justifies a simple picture of PL in a dilute 2DEG, according to 
which emission occurs from isolated, well-defined bound complexes 
($X$ and $X^-$'s), and hence it is virtually insensitive to $\nu$.
In particular, this explains the absence of an $X^-_{\rm td}$ peak 
even in the PL spectra \cite{Hayne99} showing strong recombination 
of a higher-energy triplet state $X^-_{\rm tb}$ (although the 
$X^-_{\rm td}$ emission has been eventually detected at very low 
temperatures \cite{Yusa01,Schuller02}).
An interesting feature in Fig.~\ref{fig05} is also merging of $\hbar
\omega_{\rm tb}$ and $\hbar\omega_{\rm td}$ which has actually also 
been observed experimentally at $\nu\approx{1\over3}$ \cite{Yusa01}.

\section{Anyon excitons}\label{sec_AX}

The fractionally charged ``anyon excitons'' have been predicted 
to form in strongly asymmetric quantum wells or heterostructures, 
in which the perpendicular electric field produced by the doping 
layer spatially separates conduction electron ($e$) and valence hole 
($v$) layers by a distance $d\sim\lambda$ \cite{Chen93,Chen94,fcx-e}.
In such situation, the $v$--$e$ attraction becomes too weak on the 
characteristic 2DEG correlation energy scale and the resolution of 
the attractive Coulomb potential of the hole becomes too low on the 
characteristic 2DEG length scale, and the 2DEG retains its original 
Laughlin correlations even in the presence of the hole injected 
optically in a PL experiment.
Unlike in symmetric structures, because of the reversed ordering of 
the $e$--$e$ and $v$--$e$ energy scales, the charge of the hole $v$ 
injected into the 2DEG is no longer screened with ``real'' electrons 
$e$, but with the fractionally charged Laughin quasielectrons (QE's) 
\cite{fcx-e} or reversed-spin quasielectrons (QE$_{\rm R}$) 
\cite{Rezayi87,qer}.

\begin{figure}
\begin{center}
\includegraphics{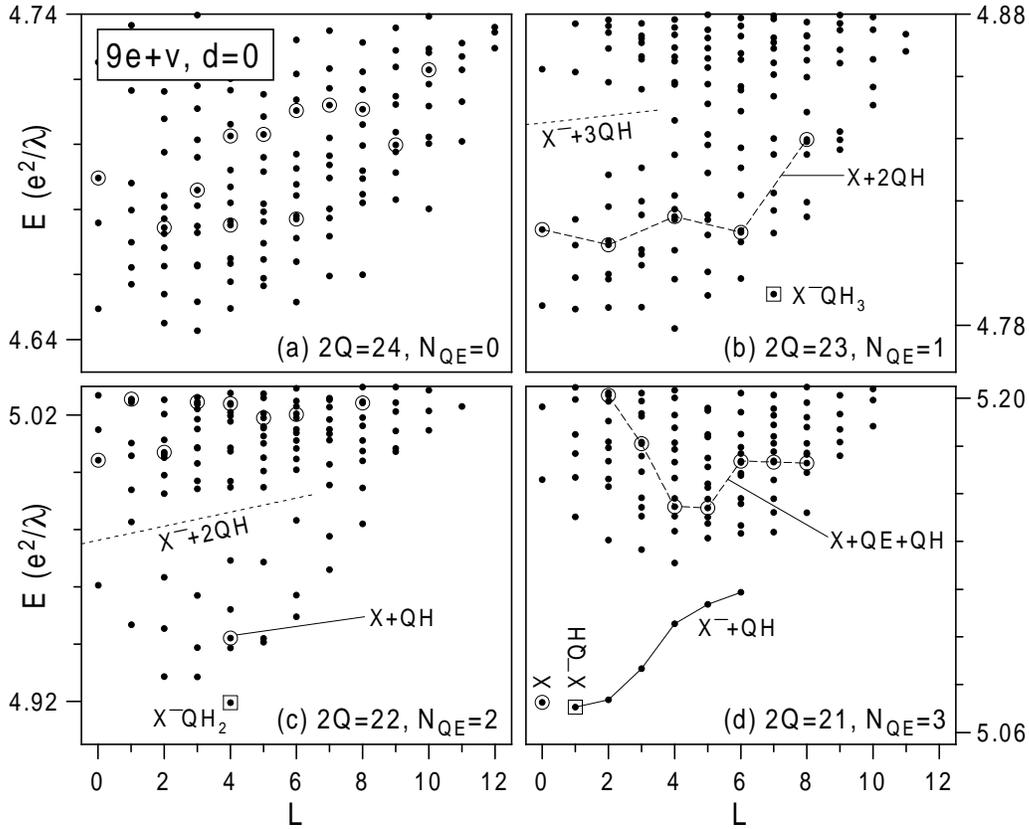}
\end{center}
\caption{
   The energy spectra (energy $E$ vs.\ angular momentum $L$) of 
   an ideal $9e+v$ system (no LL mixing and zero quantum well width) 
   calculated on Haldane sphere with LL degeneracy $2Q+1=24$ (a), 
   23 (b), 22 (c), and 21 (d).
   The $e$--$v$ layer separation is $d=0$.}
\label{fig06}
\end{figure}

\begin{figure}
\begin{center}
\includegraphics{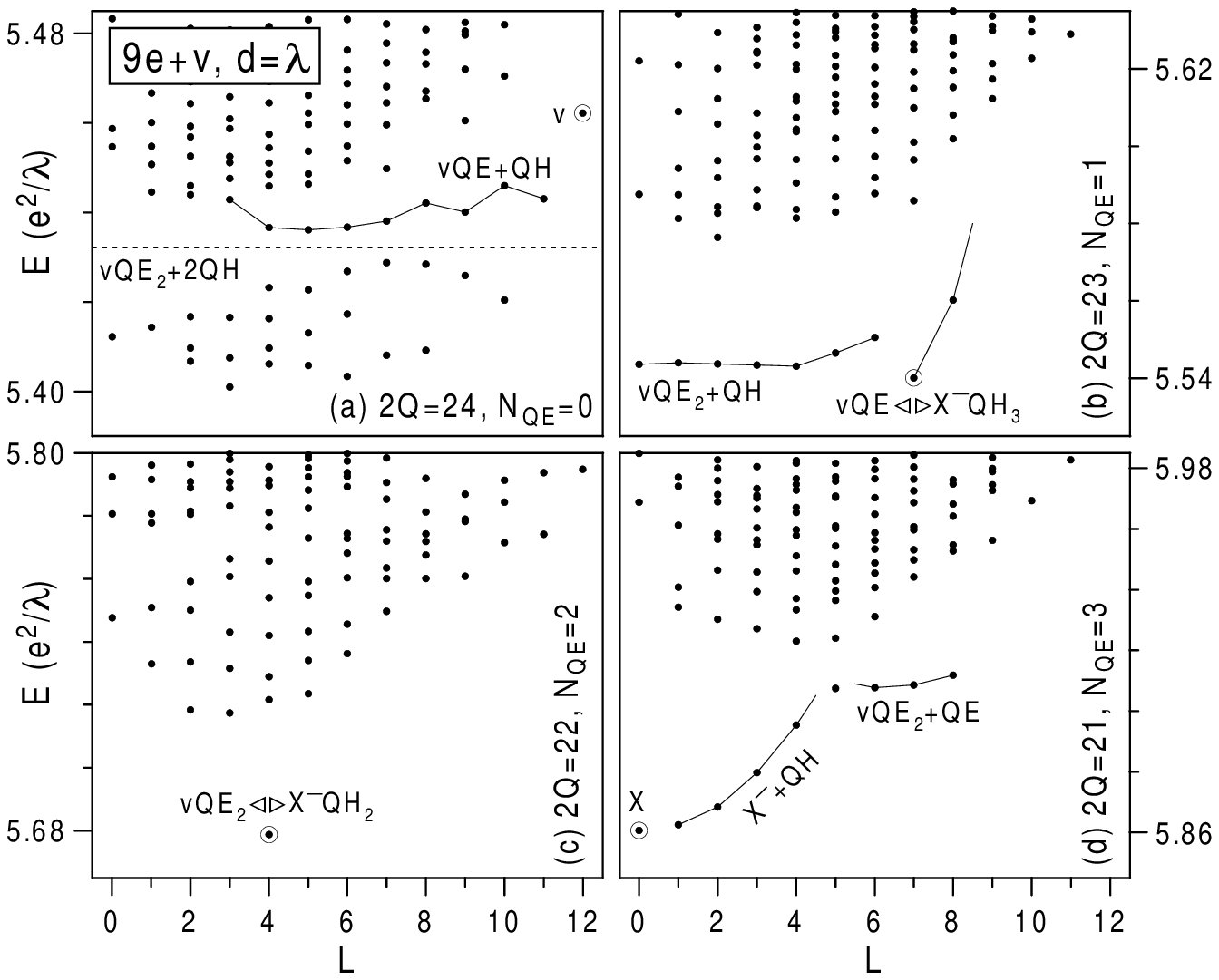}
\end{center}
\caption{
   The same as in Fig.~\ref{fig06} but for $d=\lambda$.}
\label{fig07}
\end{figure}

\begin{figure}
\begin{center}
\includegraphics{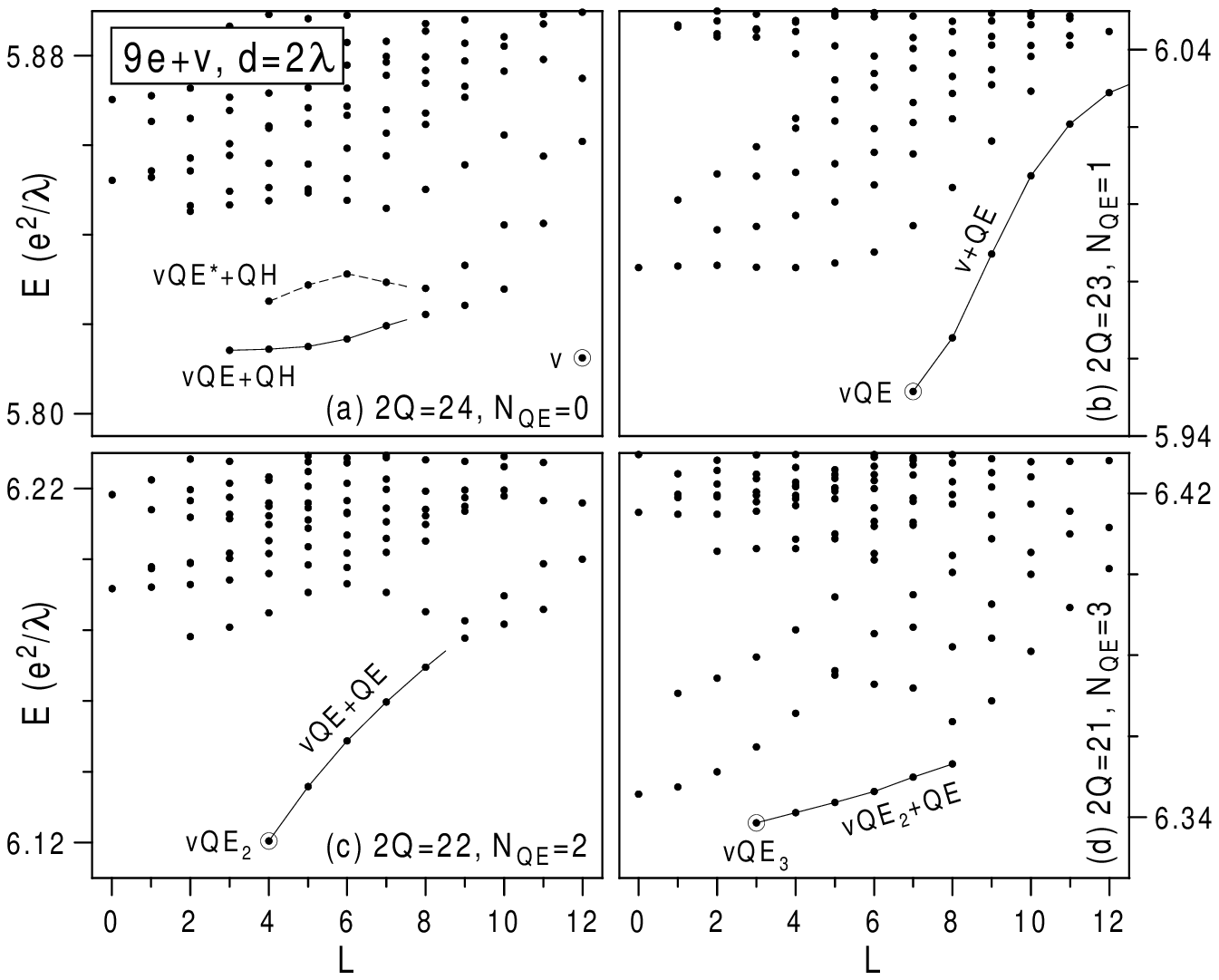}
\end{center}
\caption{
   The same as in Fig.~\ref{fig06} but for $d=2\lambda$.}
\label{fig08}
\end{figure}

The energy spectra of $9e+v$ systems at different values of the 
monopole strength $2Q$ corresponding to $N_{\rm QE}=1$, 2, and 3 
QE's in the Laughlin $\nu={1\over3}$ state of 9 electrons 
interacting with the hole have been shown in Figs.~\ref{fig06},
\ref{fig07}, and \ref{fig08} for different values of the $v$--$e$ 
layer separation, $d=0$, $\lambda$, and $2\lambda$.
These spectra have been calculated for a very ideal situation,
without taking into account the LL mixing or finite well width,
so $d/\lambda$ must be regarded as an effective parameter 
controlling the strength and resolution of the perturbation
potential introduced in the 2DEG by the presence of the hole, 
rather than as an actual displacement of $e$ and $v$ wave 
functions in an experimental system \cite{fcx-e}.

In Fig.~\ref{fig06} ($d=0$; the ``strong coupling'' regime), 
the $X^-_{\rm td}$, which is the only bound $X^-$ state in the 
lowest LL, is the most stable quasiparticle, and the anyon excitons
do not form.
The open circles mark the so-called ``multiplicative'' states in 
which the $L=0$ exciton decouples from the remaining 8 electrons
due to the ``hidden'' symmetry (the exact $e$--$v$ particle--hole 
symmetry in the lowest LL) \cite{MacDonald90}.
All other, non-multiplicative low-energy $9e+v$ states contain
an $X^-$ interacting with the remaining 7 electrons.
These states can be well described within the generalized composite 
fermion model \cite{x-cf} for the two-component ($7e+X^-$) Laughlin 
liquid.
Depending on the value of $2Q$ that varies between 24 and 21, the 
lowest-energy $7e+X^-$ states contain between zero and three quasiholes 
(QH's) analogous to Laughlin quasiholes of a one-component electron 
liquid.
The residual QH--$X^-$ attraction whose pseudopotential can be
extracted from the $X^-+{\rm QH}$ band marked in frame (d), 
leads to the formation of $X^-{\rm QH}$ and $X^-{\rm QH}_2$ very 
weakly bound states and of an excited (unstable) $X^-{\rm QH}_3$ 
states, identified in frames (b), (c), and (d), respectively.

In Fig.~\ref{fig07} ($d=\lambda$; intermediate-coupling regime), 
new low-energy bands of states emerge in addition to those containing
the $X$ or $X^-$'s.
We interpret these new states as the anyon exciton states.
In some cases the two type of states occur in the same spectrum.
For example, the $v$QE$_2$--QE band in Fig.~\ref{fig07}(c) coexists 
with the $X$ state and the $X^-$--QH band.
In other cases, the low-lying $X$ or $X^-$ states occur at the same 
$L$ as the low-lying anyon exciton states, and the transition between 
the two is continuous.
For example, $v$QE$_2$ is mixed with $X^-$QH$_2$ in Fig.~\ref{fig07}(b),
and $v$QE is mixed with $X^-$QH$_3$ in Fig.~\ref{fig07}(a).

In Fig.~\ref{fig08} ($d=2\lambda$; weak-coupling regime), well 
developed anyon exciton bands occur.
The isolated $v$QE, $v$QE$_2$, and $v$QE$_3$ states are the ground 
states in the spectra corresponding to $N_{\rm QE}=1$, 2, and 3, 
respectively.
Their angular momenta $l_{\rm AX}$ are obtained by adding $l_h=Q$ and 
$l_{\rm QE}=Q^*+1$, where $2Q^*=2Q-2(N-1)$ is the effective monopole 
strength in the composite fermion picture\cite{Jain89,parent} and 
$2Q=3(N-1)-N_{\rm QE}$.
Similarly, the angular momenta of states containing an anyon exciton 
and the excess QP's result from adding $l_{\rm AX}$ and $l_{\rm QP}$.

\begin{figure}
\begin{center}
\includegraphics{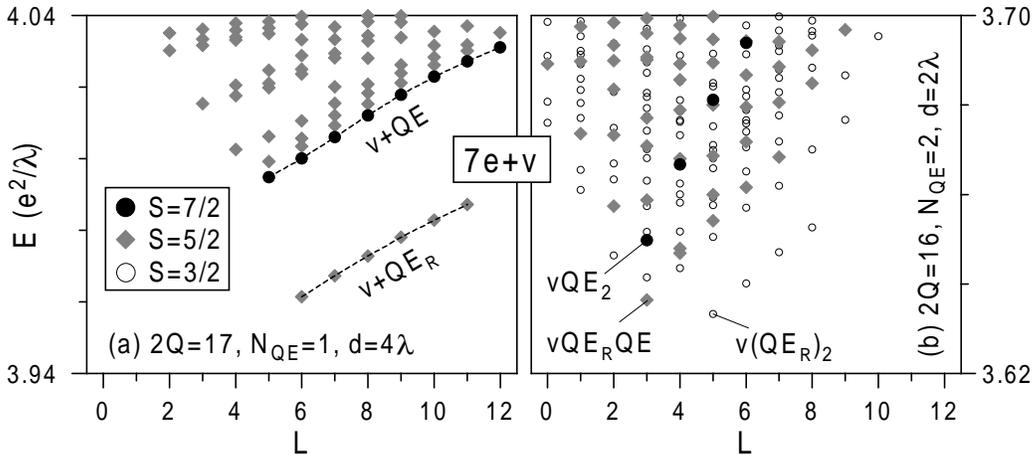}
\end{center}
\caption{
   Similar to Fig.~\ref{fig06} but for $7e+v$ system
   at LL degeneracy $2Q+1=17$ (a) and 16 (b), and including 
   electron spin excitations.
   The $e$--$v$ layer separation is $d=4\lambda$ (a) and 
   $2\lambda$ (b).}
\label{fig09}
\end{figure}

In Fig.~\ref{fig09} we show similar spectra for the $7e+v$ system, 
but now including the possible electron spin excitations \cite{qer}.
In addition to the spinless anyon excitons $v$QE and $v$QE$_2$,
the ``reversed-spin anyon excitons'' $v{\rm QE}_{\rm R}$, and 
$v{\rm QE}_{\rm R}{\rm QE}$, and $v({\rm QE}_{\rm R})_2$ can be
identified, in which one or more QE's are replaced by the 
reversed-spin quasielectrons, QE$_{\rm R}$'s.

Similarly as it was for $X^-$'s, the translational symmetry of 
an isolated anyon exciton leads to the conservation of its $L$ 
and $L_z$ in the emission process.
This leads to the strict optical selection rules that can only be 
broken by collisions or disorder.
The recombination of an anyon exciton state formed in a Laughlin 
$\nu=(2p+1)^{-1}$ electron liquid occurs through annihilation of 
a well defined number of QE's and creation of an appropriate number 
of QH's \cite{Chen93,Chen94}.
It turns out that the processes involving more than the minimum 
number of QP's all have negligible intensity, which for $p=1$ 
($\nu={1\over3}$) leaves only the following four possible 
recombination events: $v+n{\rm QE}\rightarrow (3-n){\rm QH}+\gamma$,
where $n=0$, 1, 2, or 3, and $\gamma$ denotes the photon.
When the angular momentum conservation law is applied to the above 
recombination events, we obtain \cite{fcx-pl} that the only radiative 
anyon excitons are $v$QE* (the first excited state of a $v$--QE pair),
$v$QE$_{\rm R}$, and $v$QE$_2$, while all others (including $v$QE) 
are ``dark.''

Because the formation of radiative anyon excitons depends on the 
presence of QE's or QE$_{\rm R}$'s in the 2DEG, the magneto-PL 
spectrum is expected to change discontinuously at $\nu={1\over3}$.
Such anomalous behavior has actually been observed experimentally
\cite{Heiman88}.

\section{Spin waves and skyrmions}\label{sec_Sky}

The integral quantum Hall system near $\nu=1$ with spin excitations 
contains a small number of reversed-spin electrons $e_{\rm R}$ and
spin holes $h$, and it is very similar to the dilute system of
conduction electrons $e$ and valence holes $v$ in the lowest LL.
The important difference is that the energy of a $k=0$ spin wave 
(which plays the role of an interband exciton) is equal to the
electron Zeeman splitting, $E_{\rm Z}$, which can be made small
compared to the characteristic interaction energy, $e^2/\lambda$.
Therefore, it is possible to achieve experimentally the situation 
in which the skyrmions (the $e_{\rm R}$--$h$ analogues of interband 
$X^-$'s) are truly stable ground states of the system \cite{Sondhi93,%
Palacios96}, with infinite lifetimes which are not limited by radiative 
recombination.

\begin{figure}
\begin{center}
\includegraphics{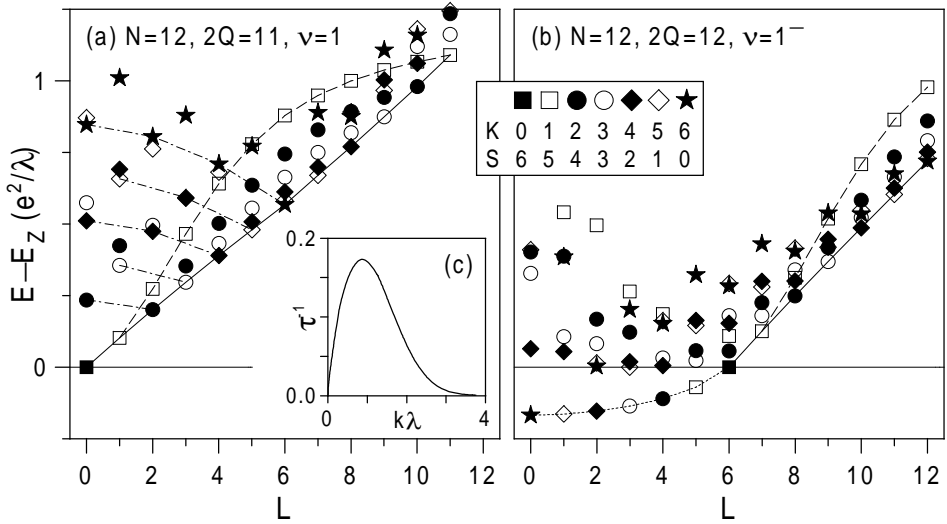}
\end{center}
\caption{
   The energy spectra (energy $E$ vs.\ angular momentum $L$) of 
   an ideal $12e$ system (no LL mixing and zero quantum well width) 
   with spin excitations in the integral quantum Hall regime,
   calculated on Haldane sphere with LL degeneracy $2Q+1=12$ (a) 
   and 13 (b).}
\label{fig10}
\end{figure}

In Fig.~\ref{fig10} we present the low energy spectra of the $\nu=1$ 
and $\nu=1^-$ (a single spin hole in $\nu=1$) states.
In this and all other spectra, only the lowest state at each $L$
and $S$ is shown and $K={1\over2}N-S$ counts the number of spin flips
away from the fully polarized ground state.

In Fig.~\ref{fig10}(a), the ground state is the ferromagnetic integral 
quantum Hall $\nu=1$ state at $L=K=0$.
Because the Zeeman energy $E_{\rm Z}$ is omitted, this state is 
degenerate with many other states with the same $L=K=0$ but with 
different values of $S_z$, and corresponding to a number $S-S_z$ 
of $k=0$ spin waves, each having energy $E_{\rm Z}=0$ and decoupled 
from one another and from the underlying $\nu=1$ state (the analogues 
of the $e$--$v$ ``mutiplicative'' states).
Remarkably, the low-energy excited states in Fig.~\ref{fig10}(a)
form a linear band with $L=K=1$, 2, \dots.
These states contain a number $K$ of spin waves each with $L=1$
and moving in the same direction so as to build up the maximum
total $L=K$.
The linear dependence of $E$ on $K$ within this band can be also
interpreted as decoupling of so correlated $L=1$ spin waves from 
one another, although different from decoupling of $k=0$ spin waves
\cite{sky}.
In particular, note that a pair of $L=1$ spin waves can be in two
states of total angular momentum $L=0$ or 2, and only the latter 
is noninteracting.

The $e_{\rm R}$--$h$ annihilation process analogous to the $e$--$v$
radiative emission can be achieved by hyperfine coupling of a 2DEG 
to localized nuclear spins. 
However, the selection rule for such process is completely different
from that governing PL.
The appropriate spectral function $\tau^{-1}(k)$ for the spin wave
creation/destruction is shown in Fig.~\ref{fig10}(c).
It has a maximum at $k\lambda\sim1$, corresponding to the 
characteristic size of the electron cyclotron orbit \cite{hyper}.

In Fig.~\ref{fig10}(b) for $\nu=1^-$ the band of states with 
$L\!=\!S\!=Q\!-\!K$ and $E\!<\!0$ appears.
These are the (anti)skyrmion states, $S_K^+=Ke_{\rm R}+(K+1)h$, 
analogous to the interband charged excitons $X_K^-$ in the 
lowest LL \cite{Sondhi93,Palacios96,sky}.
These states are not only truly long-lived (provided that $E_{\rm Z}$
can be made sufficiently small, e.g., by application of pressure 
or appropriate doping), but unlike the $X^-_K$ states they are 
connected with one another through a sequence of spin-flip 
transitions induced by the hyperfine interaction with a nuclear spin
\cite{hyper}.
The $S_K^-\leftrightarrow S_{K+1}^-$ spin-flip transitions are 
the analogues of the photon emission for the $X^-_K$ states.
However, the different selection rule described by the spectral 
function $\tau^{-1}(k)$ the inset (instead of a strict $k=0$ rule 
for the interband emission) allows these transitions in contrast
to the forbidden PL of the $X^-_K$ states in the lowest LL.
Actually, the $S_K^-\leftrightarrow S_{K+1}^-$ process is believed
to be largely responsible for the nuclear spin relaxation in
quantum Hall systems.

\begin{figure}
\begin{center}
\includegraphics{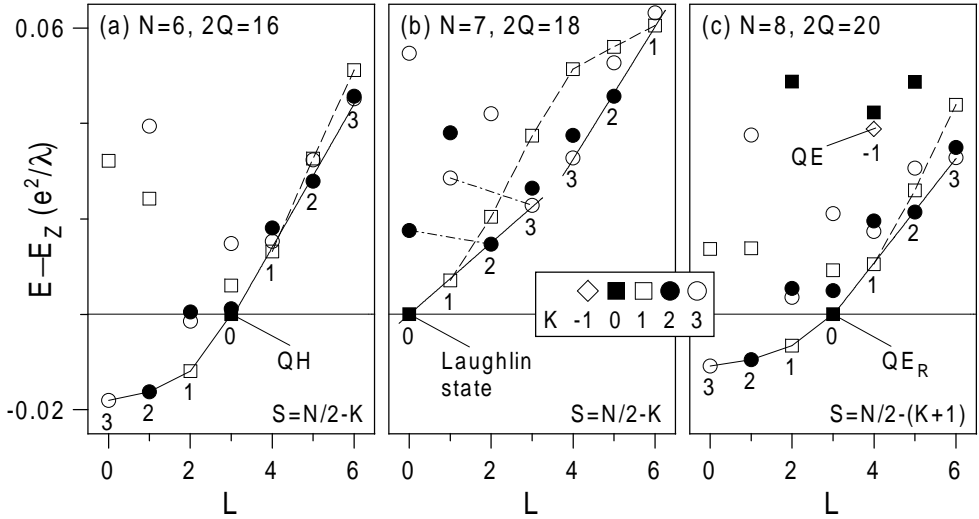}
\end{center}
\caption{
   Similar to Fig.~\ref{fig10} but for the fractional 
   quantum Hall regime ($\nu\approx{1\over3}$).}
\label{fig11}
\end{figure}

In Fig.~\ref{fig11} we show similar spectra to Fig.~\ref{fig10},
but for the fractional quantum Hall regime, near $\nu={1\over3}$.
Again, despite different character of the constituents -- elementary 
charge excitations (QE$_{\rm R}$ and QH replacing $e_{\rm R}$ and 
$h$), the same type of bound excitonic complexes are identified.
These are spin waves ${\rm QE}_{\rm R}+{\rm QH}$, skyrmions $S^-_K
=(K+1)\,{\rm QE}_{\rm R}+K\,{\rm QH}$, and antiskyrmions $S^+_K=
K\,{\rm QE}_{\rm R}+(K+1)\,{\rm QH}$.

\section{Skyrmion excitons}\label{sec_SkyX}

When a valence hole $v$ is introduced into a quantum Hall system 
with a small value of $E_{\rm Z}$, it seems possible that it might 
substitute for one of the spin holes $h$ in a skyrmion or 
antiskyrmion bound state to form yet another type of excitonic 
complexes, a skyrmion exciton \cite{Cooper97,hsky}.
Such a complex shares the properties of both pure interband and 
pure spin excitonic complexes, and for example it might both 
recombine radiatively via photon emission and couple to nuclear
spins via hyperfine interaction.
It also has a richer energy spectrum as the two kinds of holes,
$h$ and $v$ become distinguishable under actual experimental 
conditions.
Unlike in a dilute $e$--$v$ system with spin excitations where 
also three kinds of particles ($e$ could have two different spins)
were involved in a $X^-_{\rm s}$ state, different orbitals of $h$ 
and $v$ holes (e.g., due to different effective masses or different 
response to the electric field) make the $e$--$h$ and $e$--$v$ 
interactions different.
This prevents the mapping of a $h$--$v$--$e_{\rm R}$ system on 
a simple two-(iso)spin $e\!\uparrow$--$e\!\downarrow$--$v$ system 
with (iso)spin-symmetric interactions.

\begin{figure}
\begin{center}
\includegraphics{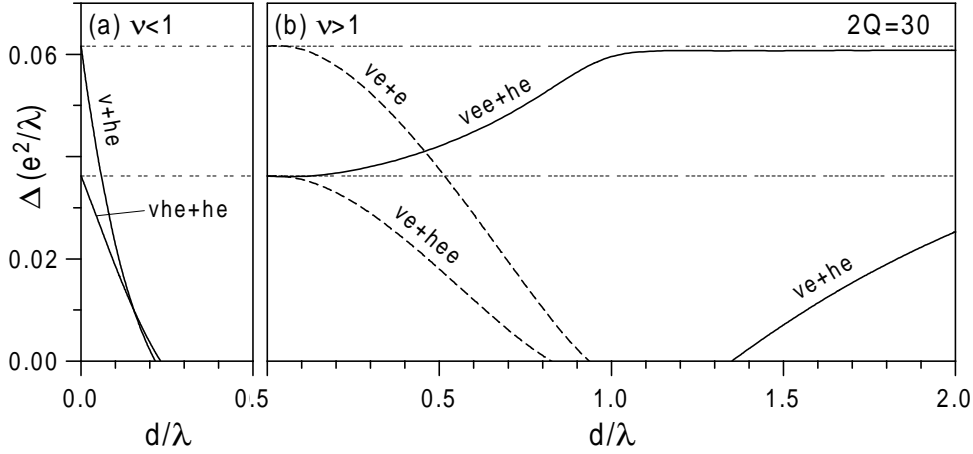}
\end{center}
\caption{
   The binding energies $\Delta$ of various skyrmion exciton states
   calculated in an ideal system (no LL mixing and zero quantum well 
   width), as a function of $eh$--$v$ layer separation $d$.
}
\label{fig12}
\end{figure}

One possible scenario for the skyrmion exciton creation might 
be the following.
When a $v$ is added to a quantum Hall state at $\nu\le1$, there are 
no negatively charged excitations it could bind.
But if $E_{\rm Z}$ is sufficiently small, $v$ may induce and bind
one or more spin waves to form a skyrmion exciton, $v\rightarrow 
vhe\rightarrow v(he)_2\rightarrow\dots$.
The binding energies of these mixed complexes are shown in 
Fig.~\ref{fig12}(a) as a function of the $eh$--$v$ layer separation 
$d$ (note that we skip subscript ``R'' in symbol $e_{\rm R}$ in this 
figure).
The situation is different and quite more complicated at $\nu>1$,
in the presence of free reversed-spin electrons or skyrmions.
Being negatively charged, they are attracted to the added hole $v$, 
and, depending on $E_{\rm Z}$, $d$, and other parameters, they can 
bind to it to form a rich variety of neutral or negatively charged 
$h$--$v$--$e_{\rm R}$ states, some of which have been indicated in 
Fig.~\ref{fig12}(b).
The fact that the binding energy for the $ve_{\rm R}+he_{\rm R}
\rightarrow vh(e_{\rm R})_2$ process remains negative for $d\le
1.35\lambda$ suggests that in symmetric structures the attraction 
between $v$ and $S_1^-=h(e_{\rm R})_2$ (or a larger skyrmion) 
causes breakup of the latter and emission of free spin waves: $v+
e_{\rm R}(he_{\rm R})_K\rightarrow ve_{\rm R}+K\times he_{\rm R}$.
This would make the equilibrium PL signal come from the same 
excitonic complex, $ve_{\rm R}$, regardless of the size of the 
skyrmions present in the 2DEG before illumination.
On the other hand, the $ve_{\rm R}$ exciton might attract a second
$e_{\rm R}$ or $S^-$ to acquire charge and become able to induce
and bind one or more spin waves.
So far these ideas have only been tested in an ideal system 
(only lowest LL included, no disorder, and zero well width), 
and more realistic calculation will be needed to verify their 
significance in actual PL experiments.

\section*{Acknowledgment}

The authors wish to acknowledge partial support from the Materials 
Research Program of Basic Energy Sciences, US Department of Energy.
AW acknowledges partial support from the Polish State Committee 
for Scientific Research (KBN) grant 2P03B05518.

\section*{References}

\end{document}